\documentclass[fleqn,usenatbib]{mnras}

\usepackage[T1]{fontenc}
\DeclareRobustCommand{\VAN}[3]{#2}
\let\VANthebibliography\thebibliography
\def\thebibliography{\DeclareRobustCommand{\VAN}[3]{##3}\VANthebibliography}
\usepackage{graphicx}	% Including figure files
\usepackage{amsmath}	% Advanced maths commands
\usepackage{amssymb}	% Extra maths symbols
\usepackage{newtxtext,newtxmath}
\title[Dynamics and habitability]{Dynamics and habitability of the TESS circumbinary systems TOI-1338 and TIC-172900988}

\author[N. Georgakarakos]{Nikolaos Georgakarakos,$^{1,2}$\thanks{E-mail: georgakarakos@hotmail.com}
\\
$^{1}$Division of Science, New York University Abu Dhabi, PO Box 129188, Saadiyat Island, Abu Dhabi, United Arab Emirates\\
$^{2}$Center for Astro, Particle and Planetary Physics (CAP3),
New York University Abu Dhabi, PO Box 129188, Saadiyat Island, Abu Dhabi, United Arab Emirates\\}

\date{Accepted XXX. Received YYY; in original form ZZZ}

\pubyear{2015}

\begin{document}
\label{firstpage}
\pagerange{\pageref{firstpage}--\pageref{lastpage}}
\maketitle

\begin{abstract}
Two circumbinary planets have been recently discovered by TESS. 
The main aim of this work is to explore whether it is possible, besides the discovered 
circumbinary planet, to have an Earth-like planet within the habitable zone of the system.  
We carry out numerical simulations over the whole range of the two habitable zones in order 
to see whether an Earth mass planet can exist there.  We find that both systems seem to be able to host 
an additional planet in their habitable zone. We construct dynamically informed habitable zones and we find that a large percentage of the habitable zone can be suitable for a planet to retain liquid water on its surface no matter what its orbital evolution will be.  Moreover, we investigate the possibility to detect an Earth-like planet in the habitable zone of the two systems. We find that for both systems, if such a planet existed, the radial velocity and astrometry signals would be rather small to be detected by our current instruments.  Some discussion is also made for the dynamical evolution of the existing planet.
\end{abstract}


%%%%%%%%%%%%%%%%%%%%%%%%%%%%%%%%%%%%%%%%%%%%%%%%%%

%%%%%%%%%%%%%%%%% BODY OF PAPER %%%%%%%%%%%%%%%%%%

\section{Introduction}
The number of discovered exoplanets keeps growing all the time. Many of these planets
are found in stellar binaries or systems of higher stellar multiplicity. Our current understanding is that a significant number of stars are found in binary or multiple stellar
systems \citep[e.g. see][and references therein]{2020Galax...8...16B}.  There are several 
pending questions regarding the formation and evolution of planets in binaries \citep{2019Galax...7...84M}.  Increasing the number of detected planets in binaries will provide valuable information to answering many of these questions.

Set in operation a few years ago, the Transiting Exoplanet Survey Satellite (TESS) has
been proven to be a valuable asset in our quest for planets outside our Solar System. So far, 
most of our recent year discoveries of circumbinary planets have been made by the {\it Kepler} mission \citep{2010Sci...327..977B,2010ApJ...713L..79K}, with the latest one being Kepler-1661b in 2020 \citep{2020AJ....159...94S}.  Last year, we had the announcement of TOI-1338b, the first circumbinary planet detected by TESS \citep{2020AJ....159..253K}.  That discovery has
been followed by another one \citep{2021arXiv210508614K}, TIC-172900988, making two the circumbinary planets found by TESS.  

Planets of the size of the Earth, not only have been an observational fact for a while now, but some of them are found in the so-called habitable zone of their host star \citep[e.g.][]{2016Natur.533..221G,2020AJ....160..117R,2020AJ....160..116G}.
The habitable zone is the region around a star where a terrestrial planet on a circular orbit can retain liquid water on its surface \citep{1993Icar..101..108K}.  In this work 
we explore the possibility of having an Earth-like planet in the habitable zone of the two recently discovered TESS circumbinary systems. By means of numerical simulations, we check whether an Earth mass planet in the habitable zone of the two systems can survive the gravitational perturbations of the stellar binary and the existing circumbinary planet
over a sufficient amount of time. We also use the concept of Dynamically Informed Habitable Zones (DIHZs hereinafter) \cite[e.g.][]{2021FrASS...8...44G} in order to evaluate the potential of that planet to support liquid water on its surface. Finally, we attempt to place some constraints on the strength of the detection signal that such a planet would create.

The structure of the paper is as follows: in section 2 we analyse the two systems from the 
dynamical evolution point of view. In section 3 we investigate the habitable zone stability of the two systems, while in section 4 we make use of the DIHZs in order to extract more information regarding the habitable zone.  In section 5 we discuss the detectability of Earth-like planets in the habitable zone of the systems, while in section 6 we provide a summary of our findings and some discussion.

\section{Dynamics of the Tess circumbinary planets}

\subsection{TOI-1338b}
TOI-1338b is a $30M_{\oplus}$ mass planet around an unequal mass binary
of a rather low orbital eccentricity. The planet revolves around the binary in 95 days with  an eccentricity of under 0.1. 
The mutual orbital inclination of the system seems to be $\sim 1^{\circ}$ \citep{2020AJ....159..253K}.
All the orbital elements and physical parameters of the system can be found in Table \ref{tab1} in the Appendix.

According to \citet{1999AJ....117..621H}, the critical semimajor axis for this system before getting unstable is 0.370 au. The semimajor axis of TOI-1338b is $a_p=0.4491 au$.  
Therefore the system appears to be fairly stable. This is also supported by the results in \citet{2013NewA...23...41G}, assuming though circular orbits. The period ratio of the system is 6.51, which implies that it is likely that our system is located in a stable area in parameter space between unstable resonances, something similar to what happens with several of the Kepler circumbinary planets \cite[e.g. see][]{2015MNRAS.446.1283C,2016AstL...42..474P}.

In \citet{2015ApJ...802...94G} we developed an analytical framework for describing the motion of nearly coplanar circumbinary planets on orbital as well as secular timescales. The method covered all binary eccentricities and it also included a post-Newtonian correction for the binary orbit. The model was getting less reliable as the planet would acquire some eccentricity ($\gtrapprox 0.2$.).

According to the analytical model, the planet is expected to cause the binary pericentre to precess at a rate of 
\begin{equation}
K_3 =\frac{3}{4}\frac{\sqrt{G}m_pa^{\frac{3}{2}}_{b}\sqrt{1-e^2_{b}}}{(m_1+m_2)^{\frac{1}{2}}a^3_{p}},
\label{aps1}
\end{equation}
where 
$G$ is the gravitational constant, $m_1$ and $m_2$ are the masses of the two stars, $m_p$ is the mass of the planet, $a$ denotes semimajor axis while $e$ denotes eccentricity. The indices $b$ and $p$ refer to the binary and planet respectively.  In addition, the binary pericentre precesses due to general relativity at a rate of (to leading order):
\begin{equation}
K_{3GR} =\frac{3G^{\frac{3}{2}}(m_1+m_2)^{\frac{3}{2}}}{c^2a_b^{\frac{5}{2}}(1-e^2_{b})}
\label{aps2}
\end{equation}
where $c$ is the speed of light in vacuum.
Hence, the rate of the binary pericentre precession is $K_3+K_{3GR}$.  For the TOI-1338 binary, the above quantity is 0.0005404 degrees over a binary period; 0.0004272 degrees from classical theory and 0.0001132 degrees from relativity. These numbers are in good agreement with \citep{2020AJ....159..253K}, where the overall best fitting model gives 0.0005715 degrees per cycle. This value includes the contribution of tides to the advancement of the argument of pericentre, which is a couple of orders smaller than the other two. 

The forced eccentricity of the circumbinary planet is given by 
\begin{equation}
e_{pf}=\frac{K_2}{K_1-K_3},
\label{forced1}
\end{equation}
with
\begin{equation}
K_1=\frac{3}{8}\frac{\sqrt{GM}m_1m_2a^2_{b}}{(m_1+m_2)^2a^{\frac{7}{2}}_{p}}(2+3e^2_{b})
\end{equation}
and
\begin{equation}
K_2=\frac{15}{64}\frac{\sqrt{GM}m_1m_2(m_1-m_2)a^3_{b}}{(m_1+m_2)^3a^{\frac{9}{2}}_{p}}e_{b}(4+3e^2_{b}).
\end{equation}
For TOI-1338b, equation (\ref{forced1}) gives $e_{pf}=0.0305$. For a circumbinary planet on an initially circular orbit, the maximum secular eccentricity will be $2e_{pf}$. If we take into account the contribution from shorter timescales, the maximum eccentricity for TOI-1338b predicted from the analytical model is 0.098, which is consistent with the 0.0928 eccentricity given by \cite{2020AJ....159..253K}. Of course, if the planet is initially on a slightly eccentric orbit, this would also affect the maximum value of the eccentricity.  The secular period of the eccentricity oscillation
and the precession period of the longitude of the pericentre is about 23.5 years which is in excellent agreement with what was found by \cite{2020AJ....159..253K}. 

\subsection{TIC-172900988}

TIC-172900988b \citep{2021arXiv210508614K} is the most recent circumbinary planet discovered by TESS.
Although the orbital parameters and mass of the planet can not be defined uniquely (six solutions with nearly equal likelihood have been found for the system - see Table \ref{tab2} in the Appendix), the planet mass is within the range $822.4 M_{\oplus} \leq m_p \leq 981 M_{\oplus}$. It's a rather big planet with at least twice the mass of Jupiter, having an orbital period of around 200 days on a nearly circular orbit. The binary itself consists of two stars with similar masses on a moderately eccentric orbit $(e_b\sim 0.45)$ of an orbital period of just under 20 days.  The highest mutual inclination among the proposed orbital solutions is $\sim 2.5^{\circ}$.  The system is similar to Kepler-34 \citep{2012Natur.481..475W} in this respect. Also, the period ratios of the two systems are similar: 10.39 for Kepler 34, while for TIC-172900988, depending on which solution we use, the period ratio ranges from 9.60 to 10.38.

The system appears to be fairly stable. The empirical criterion of \cite{1999AJ....117..621H} yields the value of 0.675 au for the critical semimajor axis. The minimum planetary semimajor axis among all solutions is 
0.86733 au.  Based on the observational data, \cite{2021arXiv210508614K} give a $0.00288^{\circ}/$ cycle for the apsidal motion of the binary. 
General Relativity accounts for $0.00017^{\circ}/$ cycle, while $0.0000232^{\circ}/$ cycle is the expected contribution due to the tidal bulges. Hence,
they conclude that the rest of the precession is because of the circumbinary planet. This hypothesis is clearly supported by the analytical model of 
\citet{2015ApJ...802...94G}.  When equation (\ref{aps2}) is evaluated for
any of the six solutions, we get a value of  $0.00017^{\circ}/$ cycle as given in \cite{2021arXiv210508614K}.  According to equation  (\ref{aps2}),
the analytical theory predicts a value between $0.0025^{\circ}$ and $0.0027^{\circ}$, depending on which solution we use, for the apsidal precession of the binary due to the presence of a planet. This seems to support the presence of a planetary body in the system.  

The forced eccentricity of the system as given by equation (\ref{forced1})
is $e_{pf}=0.0015-0.0016$.  This low forced eccentricity is to be expected as, despite the moderate binary eccentricity, the difference in the stellar masses is about $3\%$ and consequently $K_2$ gets small in (\ref{forced1}).
Therefore, we do not expect the planet to exhibit significant changes in its orbital elements on secular timescales. If we include the gravitational effects on shorter timescales, the analytical model gives a value of 0.0510-0.0580 for the maximum planetary eccentricity when the planet is on an initially circular orbit.  That value range is consistent with the planetary eccentricity of a couple of the proposed solutions as seen in Table (\ref{tab2}). The rest of the solutions have a higher $e_p$ which may be matched if we assume non-zero initial eccentricity. 

\section{Habitable zone stability}
In this section we investigate the dynamical stability of the
habitable zone of the two TESS binary systems that have been found to host a circumbinary planet.  The aim is to check whether an Earth mass 
body can survive the gravitational perturbations of the stellar 
binary and the existing planet inside the habitable zone of the system.

\subsection{Method}
\begin{figure}
\begin{center}
\includegraphics[width=85mm,height=60mm]{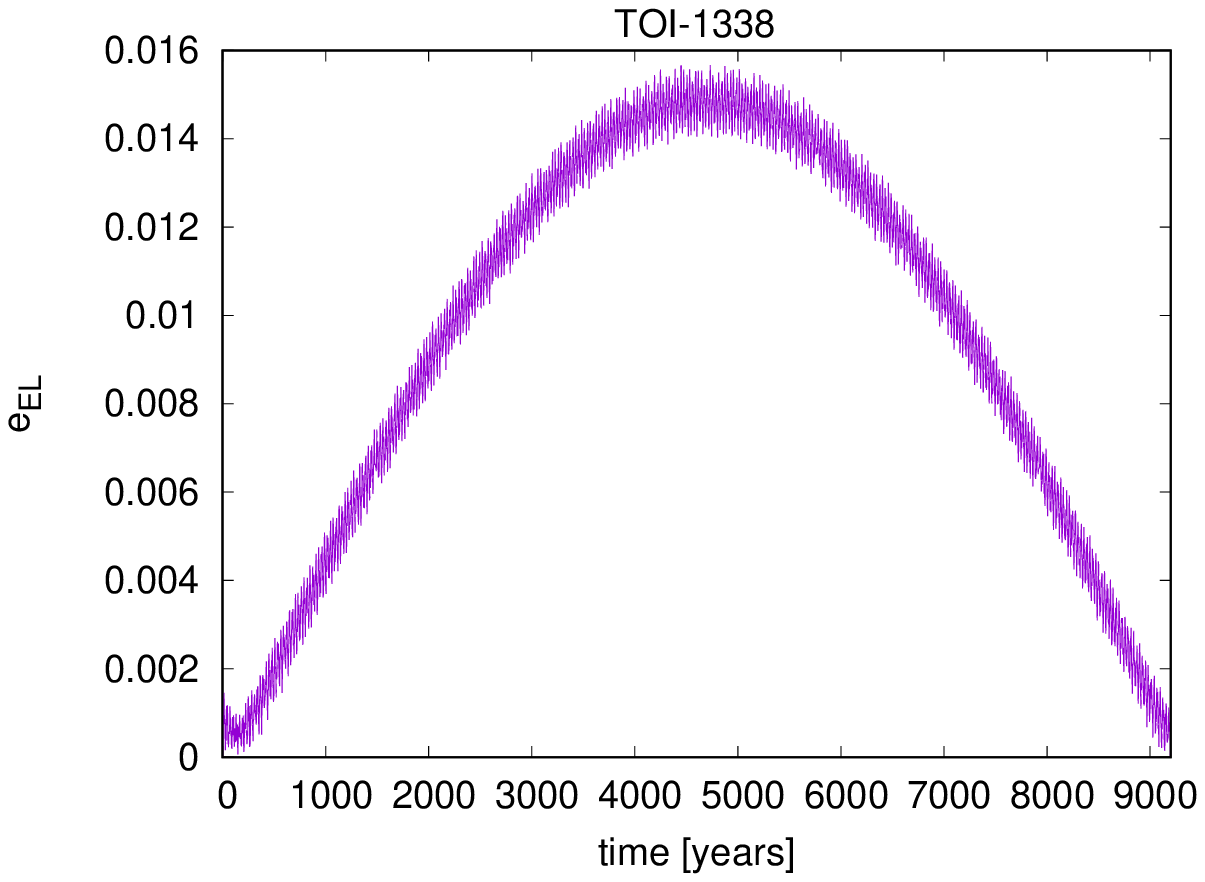}
\includegraphics[width=85mm,height=60mm]{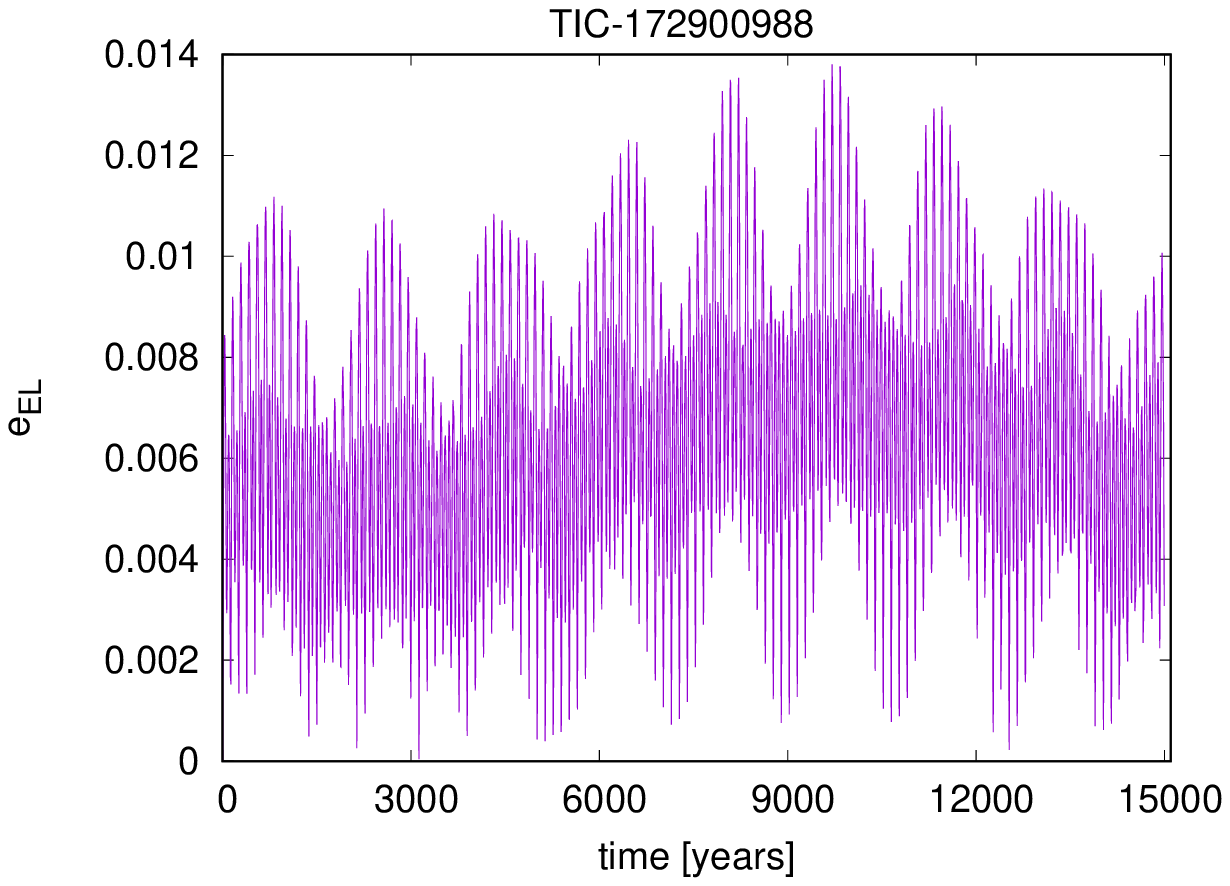}
\caption{Eccentricity evolution for an Earth mass planet at the outer edge of the habitable zone.  For TOI-1338 that location is at 2.31 au, while for TIC-172900988 is at 3.4 au.}
\label{fig1}
\end{center}
\end{figure}

In order to achieve this we ran a series of numerical experiments.
We placed an Earth mass planet within the habitable zone of each system 
and integrated numerically the full equations of motion for a time span that
was considered adequate for all sort of dynamical effects (e.g. resonances) to occur and have a potential effect on the stability of the system. For TOI-1338
that time was set to $1$ Myrs, while for TIC-172900988 the integration time was $1.5$ Myrs.  Some test numerical simulations revealed that the long period for the evolution of Earth like objects at the outer edge of the habitable zone was 
around 10000 yrs for TOI-1338 and around 15000 yrs for TIC-172900988 (see Fig. 
\ref{fig1}). Hence we chose our integration time to be around 100 times longer than those time intervals. It is clear that those long periods get shorter as we
move towards the inner edge of the habitable zone.

The Earth like bodies were placed along the habitable zone at one hundred and one equally spaced locations (including the borders of the habitable zone).  For the TOI-1338 system the step was 0.00992 au and for 
TIC-172900988 the step was 0.0146 au.  At every location the stability of the planet was tested for eight different values of its mean anomaly ($0^{\circ}-270^{\circ}$ with a stepsize of $45^{\circ}$).  A planetary orbit was classified as unstable if either its eccentricity became equal or greater than 1 or the semimajor axis recorded a value of at least $10 au$. If any of the above criteria was satisfied for at least one mean anomaly value at a given semimajor axis, then the system was classified as unstable for that semimajor axis value.  The Earth-like planet always started on an circular orbit. All bodies were on the same plane of motion.

For our numerical experiments, we used the Gauss-Radau integrator from
\cite{2010LNP...790..431E}. Regarding TOI-1338, we simulated the best fit solution as given in \cite{2020AJ....159..253K}.  In addition, we checked 
the stability of the best fit solution but with $m_p=50.5 M_{\oplus}$ which
was the best fit solution for the planetary mass plus the $1\sigma$ error. 
The errors for the rest of the parameters were such that they were not
expected to make any difference in the outcome of the simulations.
For TIC-172900988 we simulated all six solutions as given in \cite{2021arXiv210508614K}.

\subsection{Results}
\subsection{TOI-1338b}
\begin{figure}
\begin{center}
\includegraphics[width=85mm,height=60mm]{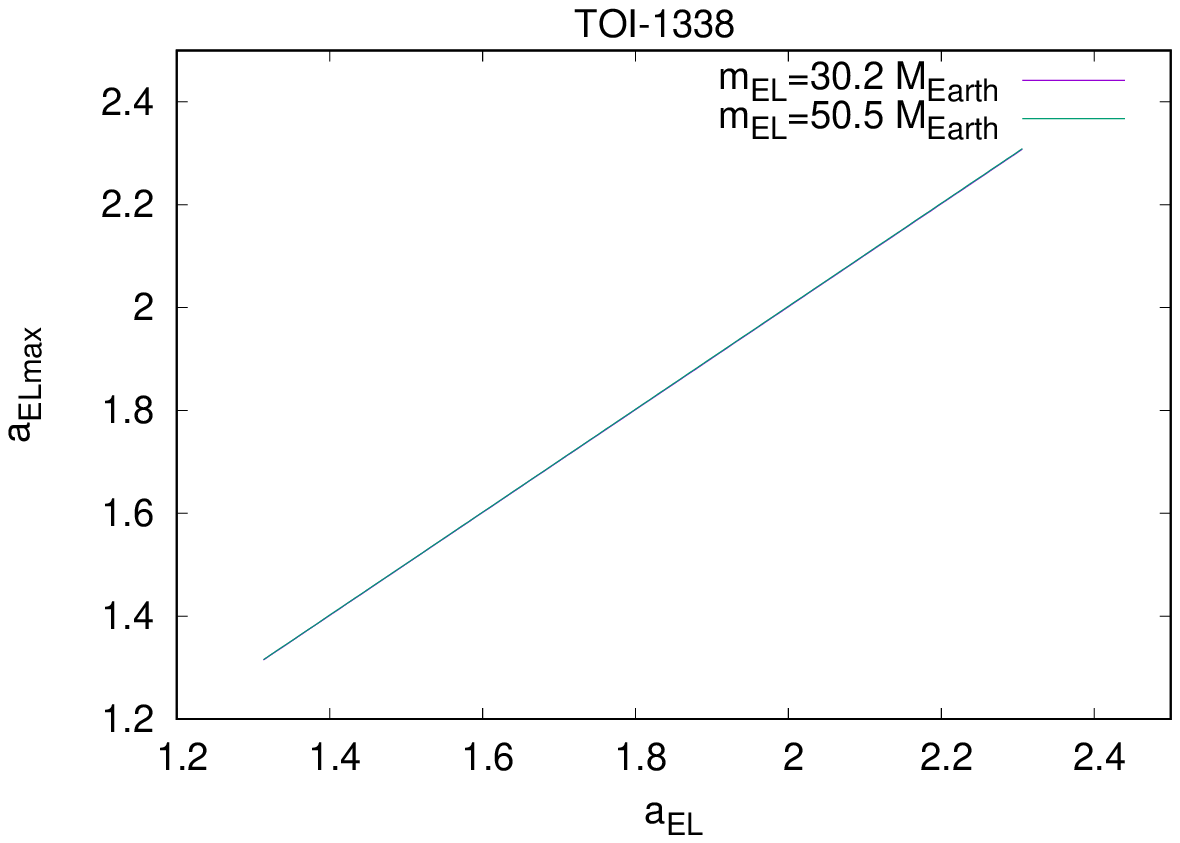}
\includegraphics[width=85mm,height=60mm]{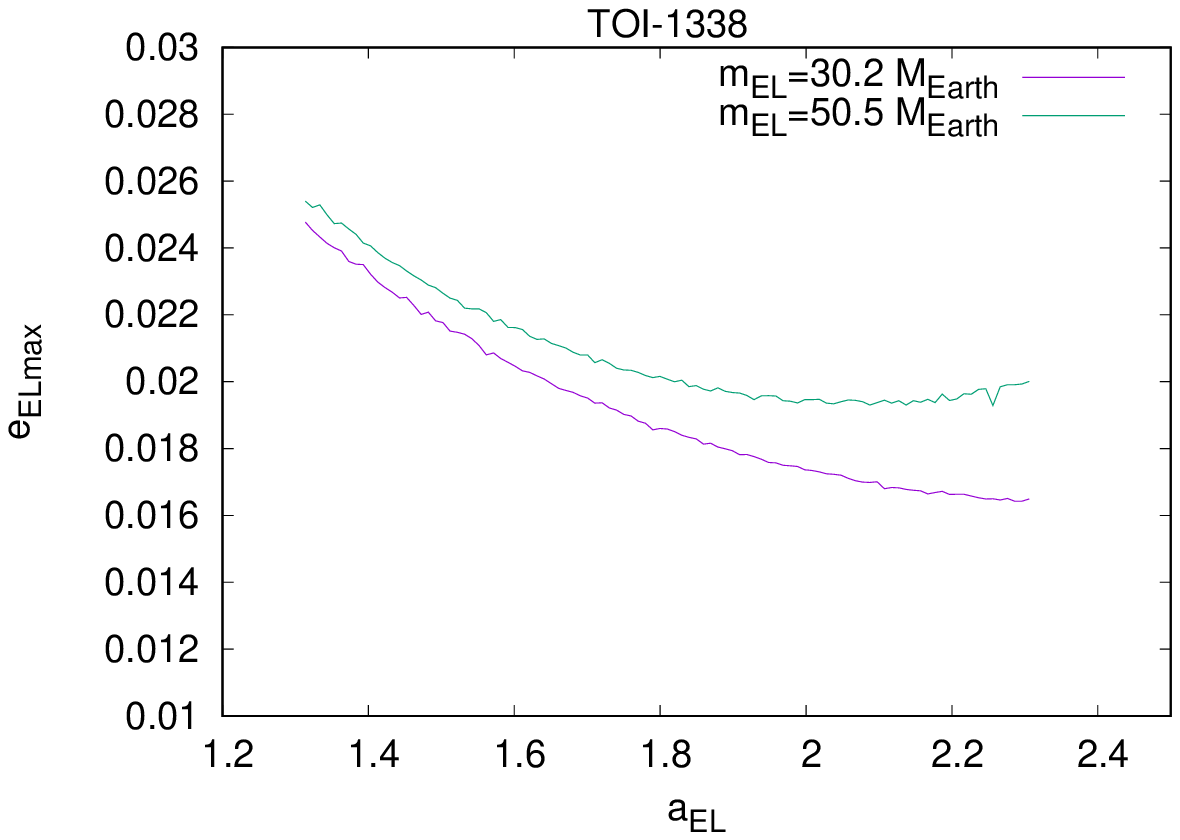}
\caption{Maximum semimajor axis and eccentricity for an Earth mass planet in the habitable zone of TOI-1338.}
\label{fig2}
\end{center}
\end{figure}

The numerical simulations for the TOI-1338 system showed no sign of instability for the fictitious Earth mass planet over the whole range of the habitable zone
 ($1.31 au - 2.31 au$). During the numerical experiments we monitored the semimajor axis and eccentricity of the Earth-like planet. The semimajor axis was almost constant.  This was the case even when the mass of TOI-1338b was raised to $m_p=50.5 M_{\oplus}$.  The eccentricity showed no significant increase over the whole length of the habitable zone.  Higher values were recorded as we moved closer to the inner edge of the habitable zone.  
Fig \ref{fig2} is a graphical representation of the above findings.  It shows the maximum semimajor axis and eccentricity recorded for the different values of the mean anomaly of the Earth-like planet at each location within the habitable zone where that body was placed. 

\subsection{TIC-172900988}

\begin{figure}
\begin{center}
\includegraphics[width=85mm,height=60mm]{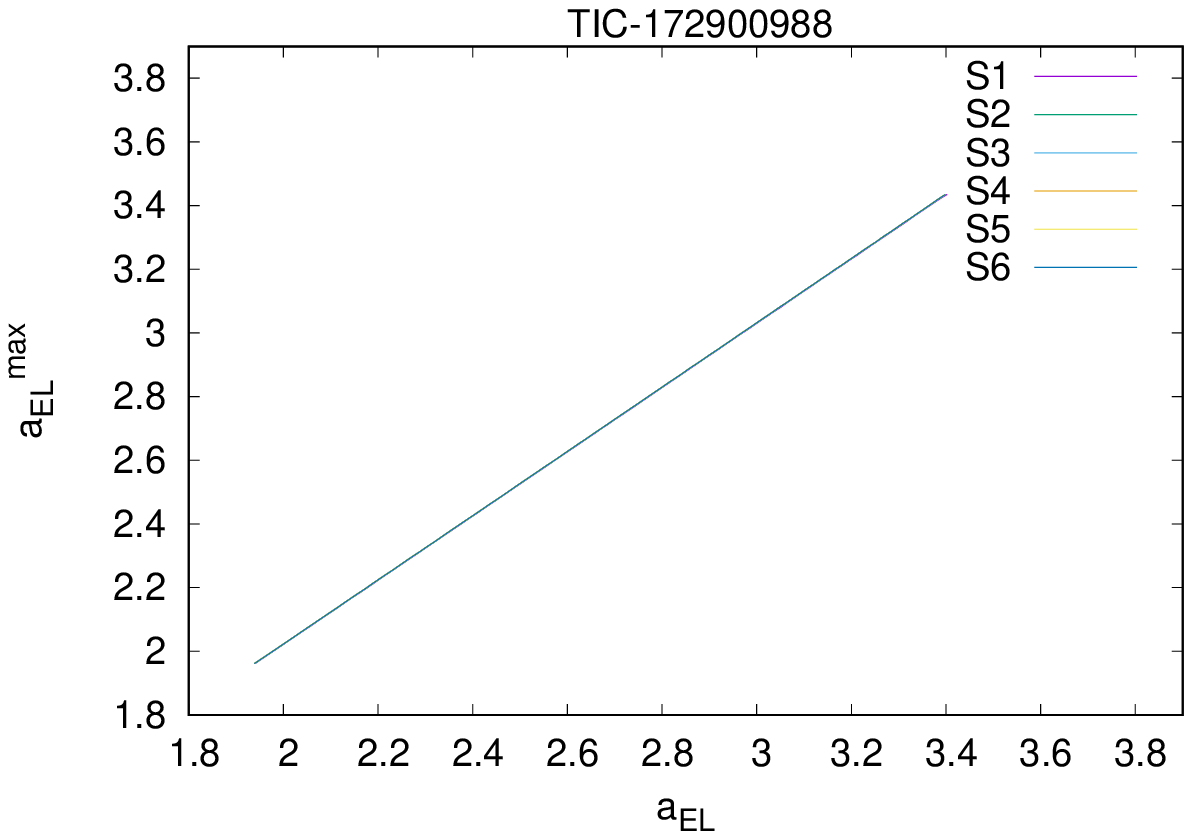}
\includegraphics[width=85mm,height=60mm]{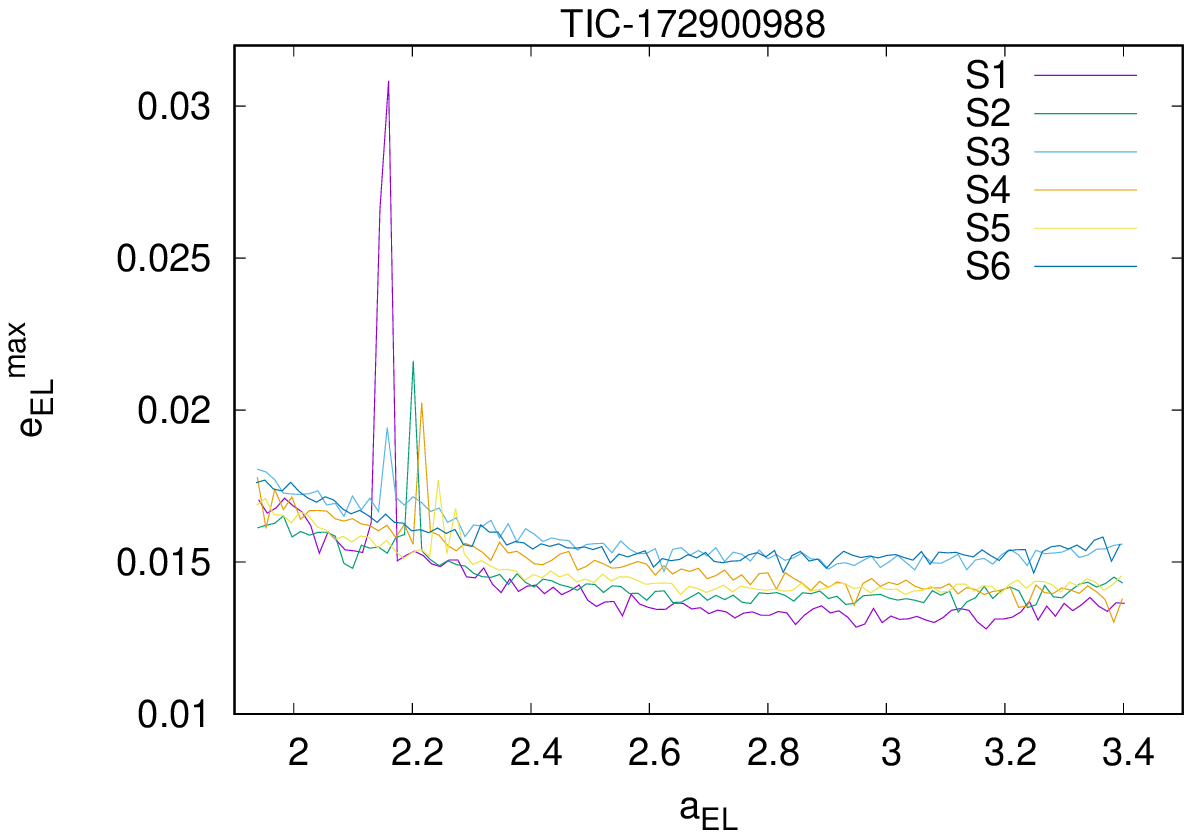}
\caption{Maximum semimajor axis and eccentricity for an Earth mass planet in the habitable zone of TIC-172900988. The spike in the maximum eccentricity is probably due
to the 4:1 resonance between the two planets.}
\label{fig3}
\end{center}
\end{figure}

The results for TIC-172900988 were similar to those of TOI-1338. For none of the six solutions 
integrated numerically the fictitious planet was unstable at any of the positions initially placed within the habitable zone ($1.94 au - 3.40 au$). The semimajor axis was almost constant,
while the maximum eccentricity values recorder were very small. Regarding the latter, similarly to what was noted for TOI-1338, higher values were found as we moved closer to the inner border of the habitable zone, but not as steeply as in that system.  
Fig \ref{fig3} is a graphical representation of the above findings.

\section {Dynamically informed habitable zones}

Dynamically Informed Habitable Zones (DIHZs) are a useful tool in the quest for habitable planets.   DIHZs are constructed in a way that takes into consideration the fact that orbital elements of planets in binary systems
vary over time due to the gravitational interactions among all bodies. 
This time dependent variation of the planetary orbit implies variation in the incoming radiation. Different planets may respond in a different ways to that time-dependent insolation variation, i.e. they may exhibit different 'climate inertia'.  This is also handled by DIHZs. 

Dynamically informed habitable zones were first introduced and used in the study of habitability of circumstellar planets in binary systems \citep{eggl-et-al-2012,2013MNRAS.428.3104E,2019MNRAS.487L..58G}.  Later on, DIHZs were used in single star systems with giant planets \citep{2018ApJ...856..155G}.
Finally, the methodology was applied to circumbinary planetary systems \citep{2018haex.bookE..61E,2020Galax...8...65E,2021FrASS...8...44G}. There are three DIHZs: the Permanently Habitable zone (PHZ), where the planet stays always within habitable insolation limits, the Averaged Habitable Zone (AHZ) where the insolation is on average within habitable values, and finally, the Extended Habitable Zone (EHZ), which is the region where the planet
stays on average plus minus one standard deviation within
habitable insolation limits. The PHZ is ideal for planets with low climate inertia. On the other hand, the AHZ assumes unlimited buffering capabilities of the planet's atmosphere, while the EHZ stands in between the other two zones.
More details about the three types of zones can be found elsewhere \citep[e.g.][]{2018ApJ...856..155G}.

We now proceed with the calculation of the habitable zones for the two TESS circumbinary systems under investigation.  In order to do that, we use the methodology described in detail in \cite{2021FrASS...8...44G}.  Using the equations
derived in that work, we obtain the limits of the classical (CHZ) and dynamically informed habitable zones for the two solutions of TOI-1338 and the six solutions for TIC-172900988. 
For systems consisting of a star and a planet on a fixed circular orbit, the limits of the CHZ are found by
\begin{equation}
\label{disthz}
r_{I,O}=\sqrt{\left(\frac{L}{S_{I,O}}\right)},
\end{equation}
where $r$ is the distance of the planet to the host star in astronomical units and $L$ is the  stellar luminosity in solar luminosities.  $S_I$ and $S_O$ 
are the effective insolation values \citep{1993Icar..101..108K}, which correspond to the number of solar constants required to trigger a runaway greenhouse process evaporating surface oceans (subscript $I$), or a snowball state freezing oceans on a global scale (subscript $O$).  They can be calculated from expressions found in \citep{2014ApJ...787L..29K}.
For a binary star, we use the following modified version of equation (\ref{disthz}) in order to calculate the borders of the CHZ \citep{2018haex.bookE..61E,2021FrASS...8...44G}:
\begin{equation}
\label{eq:disthzbin}
r_{I,O}=\sqrt{\left(\frac{L_1}{S1_{I,O}}+\frac{L_2}{S2_{I,O}}\right)},
\end{equation}
where the indices 1 and 2 refer to the two stars.  The values of all zones can be found in Table \ref{tabhz}.  For calculating the borders of the DIHZs we used maximum eccentricity and averaged squared eccentricity values obtained from the stability simulations of the Earth-like planet.

As we saw in the previous section, the entire classical habitable
zone is essentially dynamically stable for both TOI-1338 and TIC-172900988.
Due to the characteristics of the two systems (masses and orbital parameters), 
the Earth-like planet does not acquire significant orbital eccentricity.
This fact is reflected in the values of the DIHZs, especially in the values
of the borders of the PHZ which depend on the maximum eccentricity. For any of the 
two TOI-1338 solutions we investigated, the length of the PHZ is around $70\%$
of the length of the classical habitable zone.  This implies that a potentially 
habitable planet with low climate inertia could remain habitable in most of the
classical habitable zone.  If the planet has some limited buffering capabilities 
when there is change in the incoming insolation, then, that planet can retain its
liquid water on its surface for an even larger chunk of the CHZ (around $88\%$ as the limits of the EHZ indicate).

TIC-172900988 also exhibits very good potential for hosting habitable worlds.
Among the six orbital and parameter solutions proposed for that system that are in line with the observations, we notice a greater variety regarding the potential of having habitable planets. Solutions 1 and 4 provide the smallest PHZ: $68\%$ and $71\%$ respectively of the CHZ can be suitable for a planet to support liquid water on its surface no matter what its orbital evolution will be. At the other end, we have solutions 5 and 6 that provide the most extensive PHZs ($\sim 90\%$).  This is due to the fact that these solutions provide the two smallest eccentricities $e_p$ for TIC-172900988b. Small eccentricities for the giant planet imply that a terrestrial planet on an initially
circular orbit would not be expected to reach high values of its orbital eccentricity \citep[e.g. see][]{2016MNRAS.461.1512G}.  This is visible in the bottom plot of Fig.\ref{fig3}. Solutions 2 and 3 yield PHZ lengths in between the other two groups of solutions.  Finally, if our potential habitable world has a moderate capability of buffering changes in the incoming radiation, then the planet can reside at an even larger part of the CHZ, ranging from $86\%$ to $95\%$.
 
\begin{figure}
\begin{center}
\includegraphics[width=85mm,height=60mm]{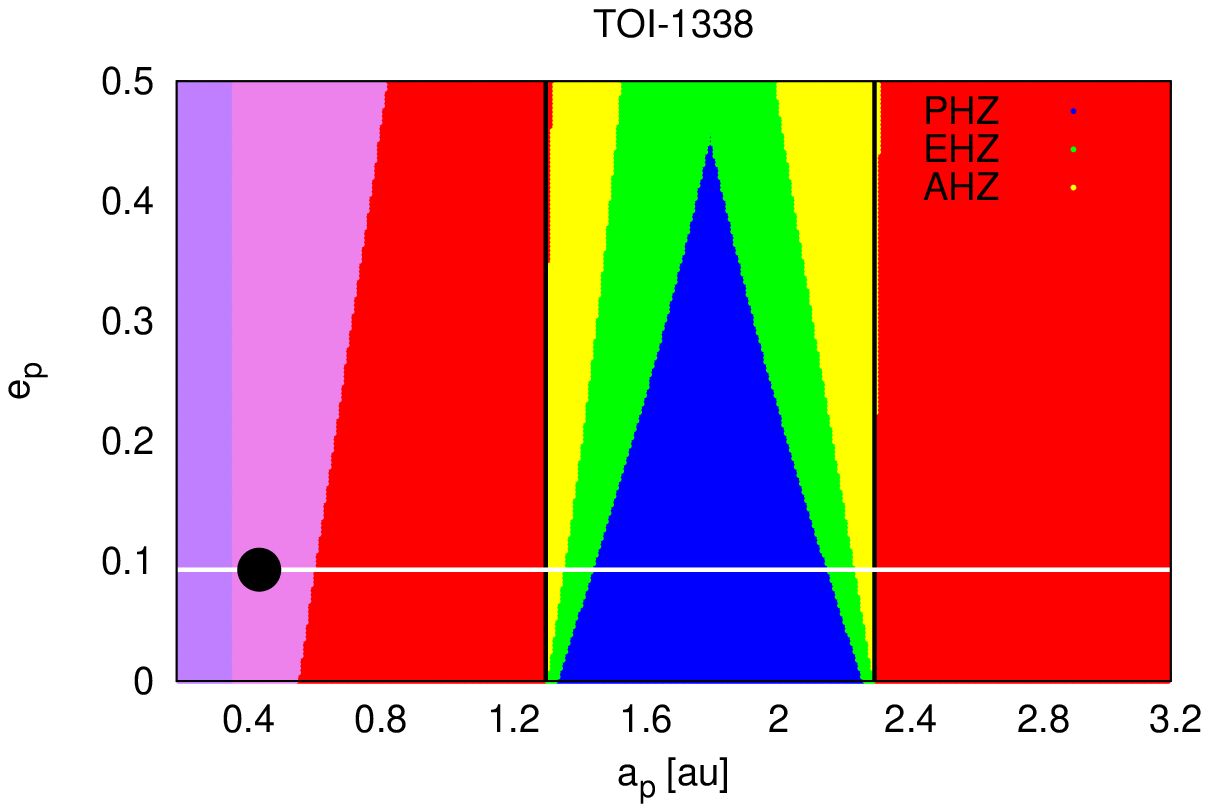}
\includegraphics[width=85mm,height=60mm]{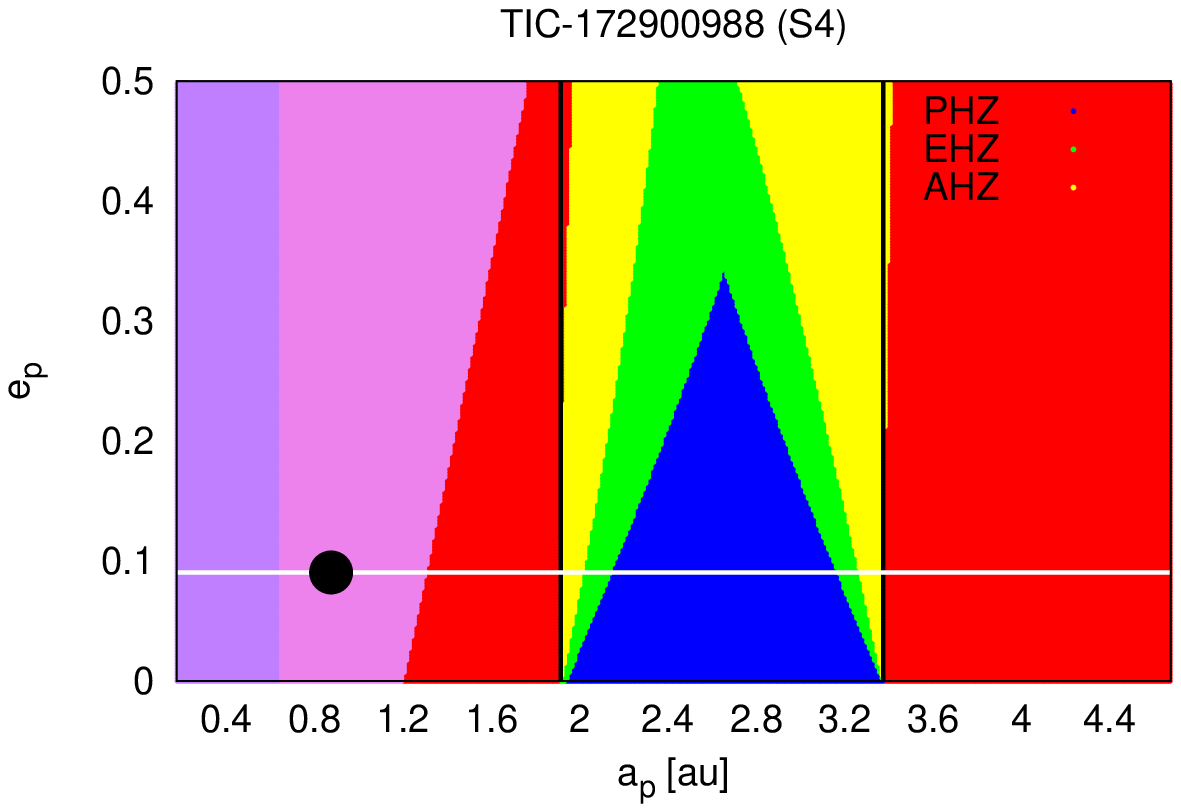}
\caption{Dynamically informed habitable zones for TOI-1338 and TIC-172900988.  The upper panel is for the case with $m_p=50.5 M_{\oplus}$, while the bottom one is for TIC-172900988 (S4). The vertical black lines are the classical zone borders. The black disc marks the current location of the existing planet and the horizontal white line is an indicator for its eccentricity value.  Blue colour: PHZ, yellow colour: AHZ, green colour: EHZ, red colour: uninhabitable, purple colour: unstable zone (binary effect - based on the empirical criterion of \protect\cite{1999AJ....117..621H}), violet colour: unstable zone (existing planet effect - based on the empirical criterion of \protect\cite{2015ApJ...808..120P}).}
\label{fig4}
\end{center}
\end{figure}

Fig. \ref{fig4} is a graphical representation of the DIHZs for the two binary systems.
In those plots we have allowed the eccentricity $e_p$ of TOI-1338b and TIC-172900988b
to vary up to much higher values so that the reader can visualize better the effect of that quantity on the length of the various habitable zones.

\begin{table*}
  \caption{Habitable zone limits for TOI-1338 and TIC-172900988.}
 \label{tabhz}
\begin{center}	
\begin{tabular}{l l l l l}\hline\\
    System & CHZ (au)& PHZ (au) & EHZ (au) & AHZ  (au) \\ \\
    \hline\\
    TOI-1338 ($m_p=30.2 M_{\oplus}$) & 1.31 - 2.31 & 1.46 - 2.16 (70\%)& 1.37 - 2.25 (88\%)& 1.31 - 2.31 (100\%) \\
    TOI-1338 ($m_p=50.5 M_{\oplus}$) & 1.31 - 2.31 & 1.45 - 2.17 (72\%)& 1.37 - 2.25 (88\%)& 1.32 - 2.30 (98\%)\\
    TIC-172900988 (S1) & 1.94 - 3.40 & 2.22 - 3.21 (68\%)& 2.03 - 3.30 (87\%)& 1.94 - 3.40 (100\%) \\
    TIC-172900988 (S2) & 1.94 - 3.40 & 2.11 - 3.25 (78\%)& 2.01 - 3.32 (90\%)& 1.94 - 3.40 (100\%)\\
    TIC-172900988 (S3) & 1.94 - 3.40 & 2.13 - 3.23 (75\%)& 2.02 - 3.31 (88\%)& 1.94 - 3.40 (100\%)\\
    TIC-172900988 (S4) & 1.94 - 3.40 & 2.16 - 3.19 (71\%)& 2.04 - 3.29 (86\%)& 1.94 - 3.40 (100\%)\\
    TIC-172900988 (S5) & 1.94 - 3.40 & 2.02 - 3.33 (90\%)& 1.97 - 3.36 (95\%)& 1.94 - 3.40 (100\%)\\
    TIC-172900988 (S6) & 1.94 - 3.40 & 2.02 - 3.33 (90\%)& 1.97 - 3.36 (95\%)& 1.94 - 3.39 (100\%)\\
       \hline
  \end{tabular}
\end{center}
\end{table*}

\section{Detectability of Earth like planets} 

We saw in the previous section that both habitable zones of TOI-1338
and TIC-172900988 are good enough to support a planet with liquid water on its surface.
A valid question is, however, whether such a planet could be detected within the range of the habitable zone. In this section we provide some estimates in order to answer the above question. 

It has been suggested that planets in binaries may be easier to detect compared to a single star system, as the gravitational interactions between the planet and the second star can boost radial velocity and astrometric signals \citep[e.g.][]{2013ApJ...764..130E}.
Following \cite{2013ApJ...764..130E} and modifying the RV semi-amplitude accordingly for a circumbinary system \citep{2019A&A...624A..68M}, the maximum value of the radial velocity achieved is
\begin{equation}
V_r^{max}=\frac{\sqrt{G}m_{EL}\sin{i}\sqrt{1+e_{EL}^{max}}}{\sqrt{a_{EL}(m_1+m_2+m_{EL})(1-e_{EL}^{max})}},
\label{rv}
\end{equation}
where $i$ is the inclination with respect to the plane of the sky.  The subscript {\it EL} refers to the Earth-like planet.

We now evaluate equation (\ref{rv}) for the TOI-1338 system over the range of its habitable zone.  For $e_{EL}^{max}$ we used the values acquired during our stability 
experiment. Also, we assume that $\sin{i}=0^{\circ}$. For both of the TOI-1338b solutions, $V_r^{max}$ does not exceed the value of $7cm/s$ at the inner edge of the habitable zone.  As we move closer to the other end of the habitable zone, that value becomes even smaller. Of course, this is something to be expected. Earth at 1 au distance from the Sun has an RV semi-amplitude of $\sim 9 cm/s$. On the other hand,  
TOI-1338 has a total stellar mass of $\sim 1.3 M_{\odot}$ and the habitable zone starts around 1.3 au. As seen from equation (\ref{rv}), the increased stellar masses and Earth-like planet semimajor axis will result in smaller values of $V_r^{max}$.
Moreover, as we saw in Fig. \ref{fig2}, the terrestrial planet's maximum eccentricity $e_{EL}^{max}$ 
remains very small throughout the habitable zone, and therefore that is another reason that $V_r^{max}$, which is an increasing function of $e_{EL}^{max}$, has low values. 

Generally, the maximum value $V_{r}$ of the radial velocity may not be the best indicator
for us in order to evaluate the potential detectability of a planet at a certain range.  This is because it is not certain that we will be able to observe the planet when it has reached its maximum orbital eccentricity or
a value close to it.  Depending on our system, the planet may exhibit a very long secular period with a significant eccentricity amplitude.  Hence, it may be more appropriate to use different indicators, such as for example the root mean square (rms) of the radial velocity measurements.  The expression has the following form \citep{2013ApJ...764..130E}:

\begin{equation}
V_r^{rms}=\frac{\sqrt{G}m_{EL}|\sin{i}|}{\sqrt{2a_{EL}(m_1+m_2+m_{EL})}}
\label{rms}
\end{equation}

\begin{figure}
\begin{center}
\includegraphics[width=80mm,height=55mm]{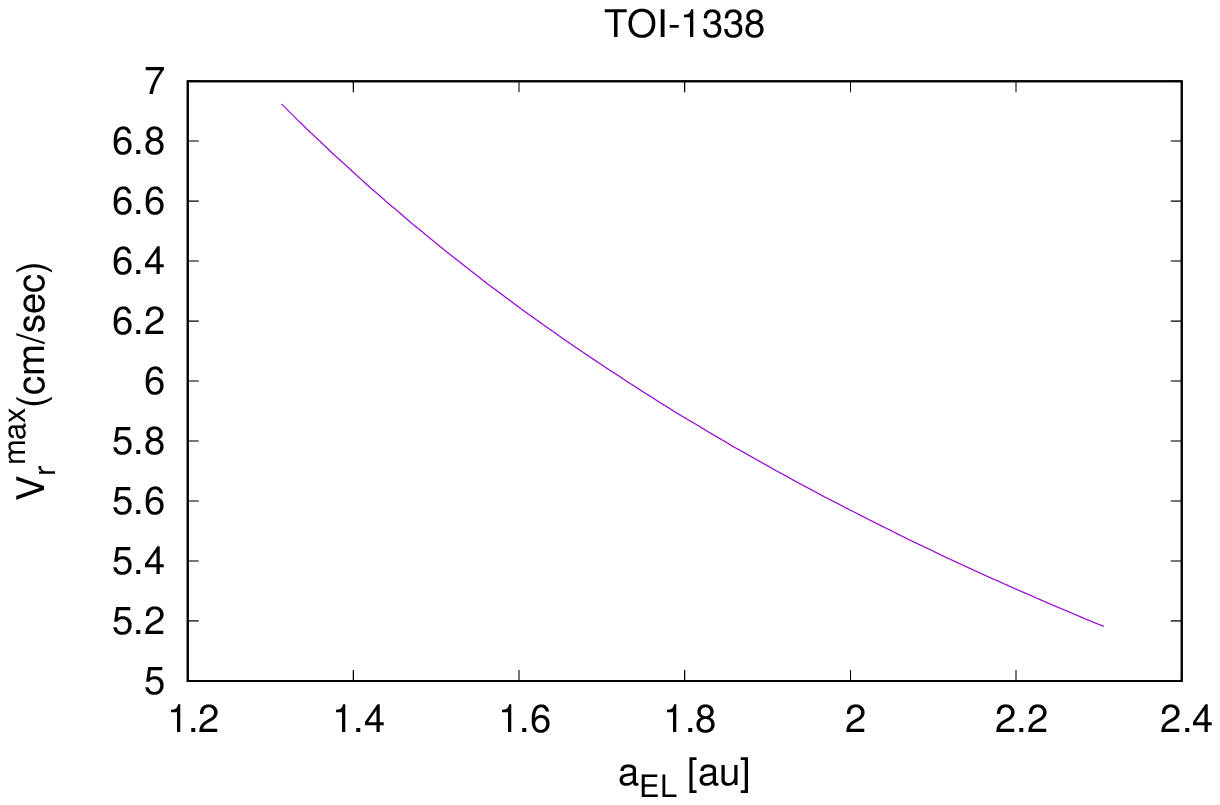}
\includegraphics[width=80mm,height=55mm]{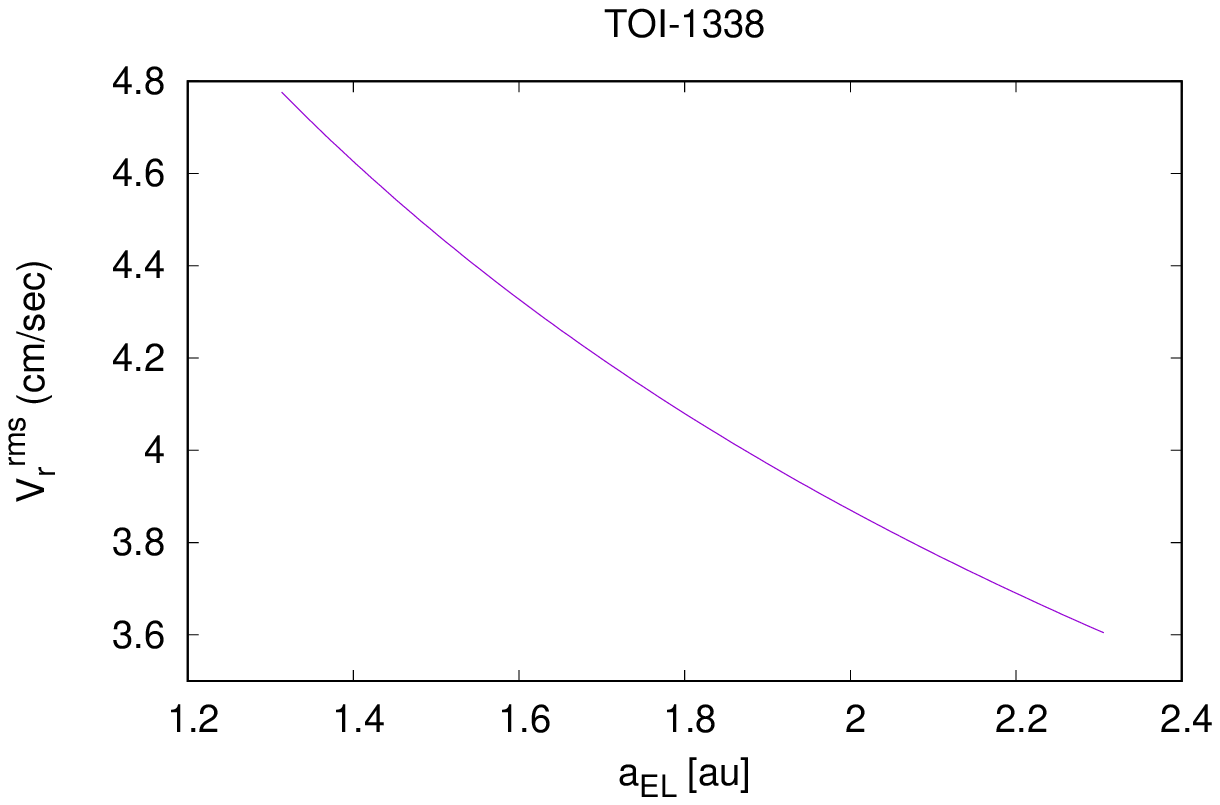}
\includegraphics[width=80mm,height=55mm]{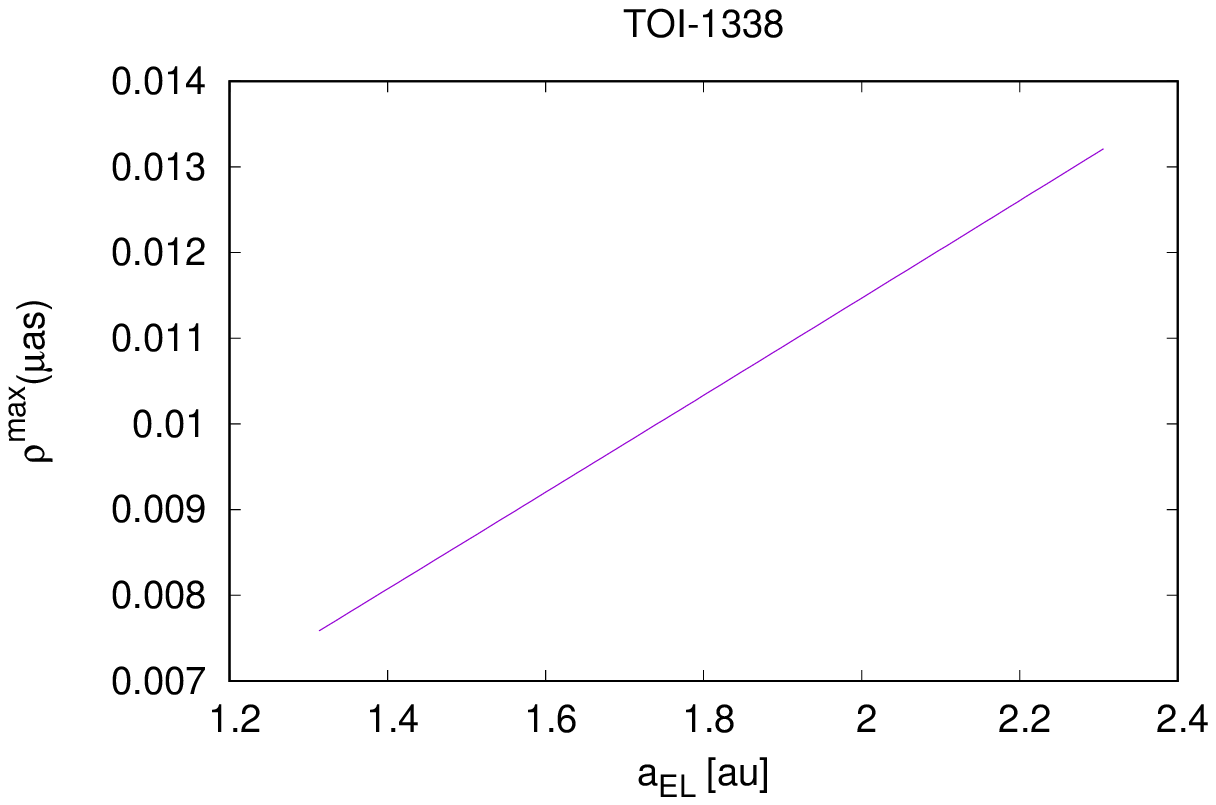}
\includegraphics[width=80mm,height=55mm]{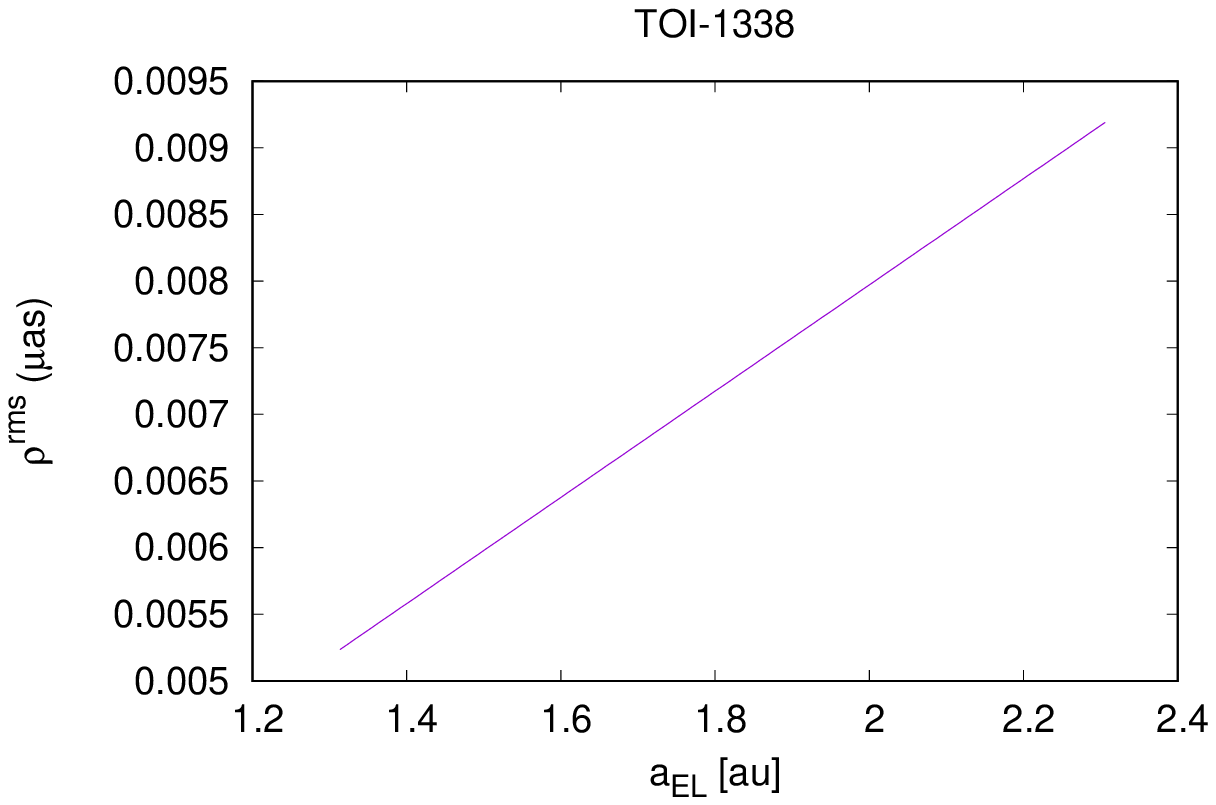}
\caption{Detectability values for the best fit solution of TOI-1338.}
\label{fig5}
\end{center}
\end{figure}

For TOI-1338, $V_r^{rms}$ is just under $4.8 cm/s$ at the inner edge of the habitable zone and it goes down as we move towards the outer border of the 
habitable zone. Fig. \ref{fig5} (upper two panels) provide the evolution of 
$V_r^{max}$ and $V_r^{rms}$ within the habitable zone for TOI-1338 best fit solution. 

If we now consider astrometry as our detection method, there are similar to the RV method expressions regarding the maximum astrometric amplitude and the astrometric rms.  These are given by \citep{2013ApJ...764..130E}:

\begin{equation}
\rho^{max}=\frac{m_{EL}a_{EL}(1+e_{EL}^{max})}{(m_1+m_2+m_{EL})d}
\label{astrom}
\end{equation}
and 
\begin{eqnarray}
\rho^{rms}&=&\frac{1}{2}\frac{m_{EL}a_{EL}}{(m_1+m_2+m_{EL})d}\left[3+\frac{9}{2}<e_{EL}^2>+\left(1+\right.\right.\nonumber\\
&+&\left.\left.\frac{3}{2}<e_{EL}^2>\right)\cos{2i}\right]^{1/2},
\label{astrorms}
\end{eqnarray}
where $d$ is the distance (in au) between the observer and the observed system and $<e_{EL}^2>$ is the averaged over mean anomaly and argument of pericentre squared eccentricity.
The reader should note here that $\rho^{max}$ is independent of the inclination $i$.

For TOI-1338, the maximum astrometric signal for an Earth mass planet in the habitable zone of the system is quite small, i.e. smaller than 0.014 $\mu as$.
The average signal is an order of magnitude lower than that. The two bottom panels of Fig. \ref{fig5} confirm that.

\begin{figure}
\begin{center}
\includegraphics[width=80mm,height=55mm]{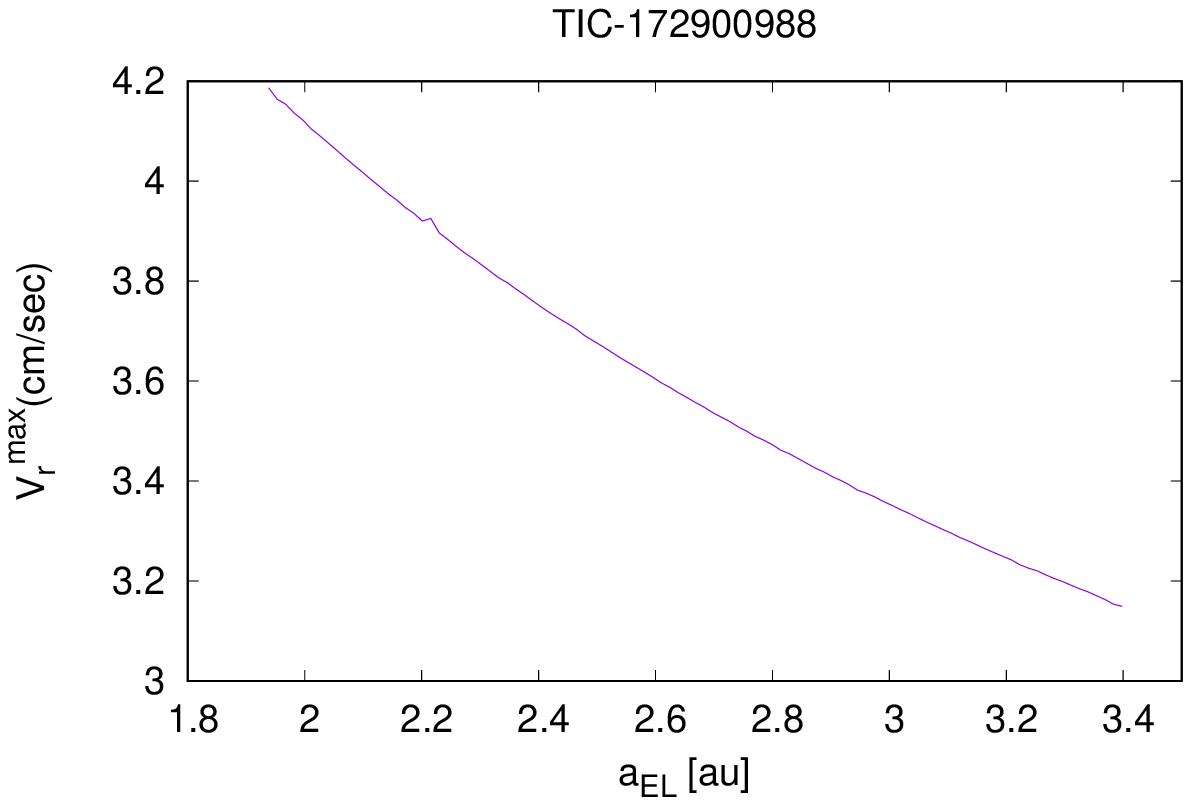}
\includegraphics[width=80mm,height=55mm]{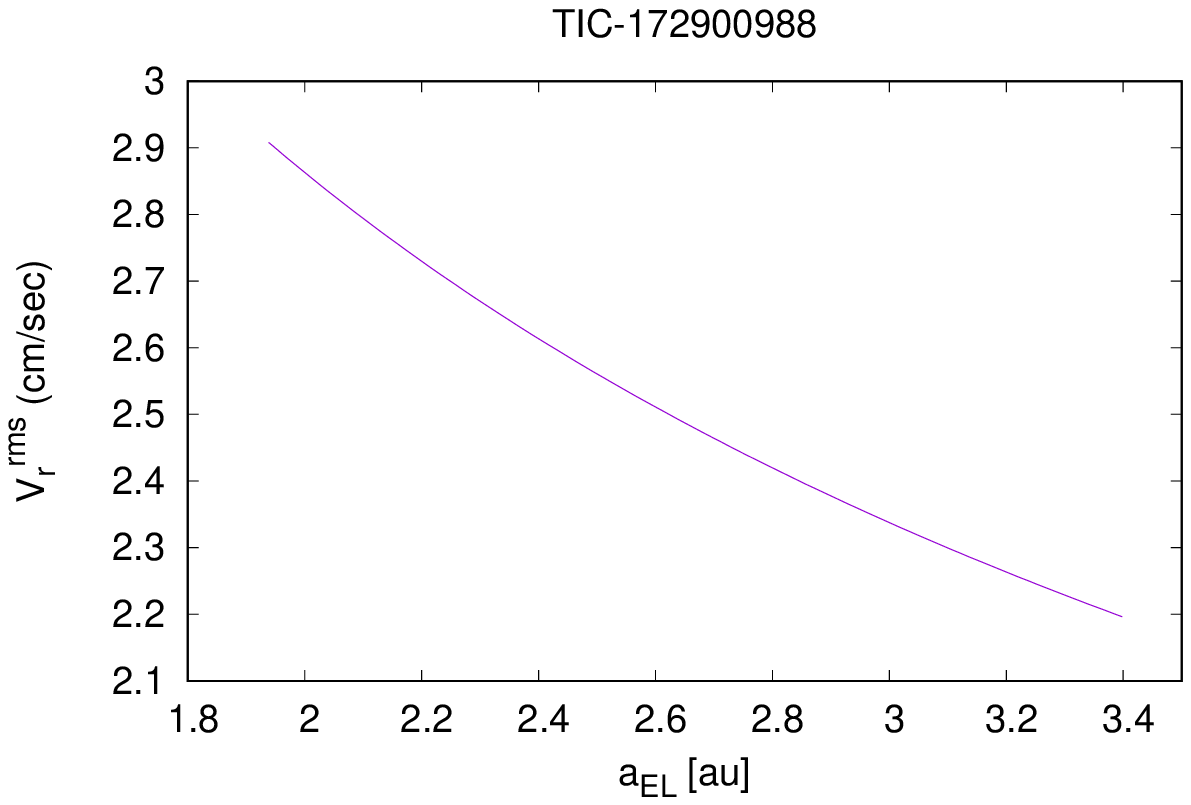}
\includegraphics[width=80mm,height=55mm]{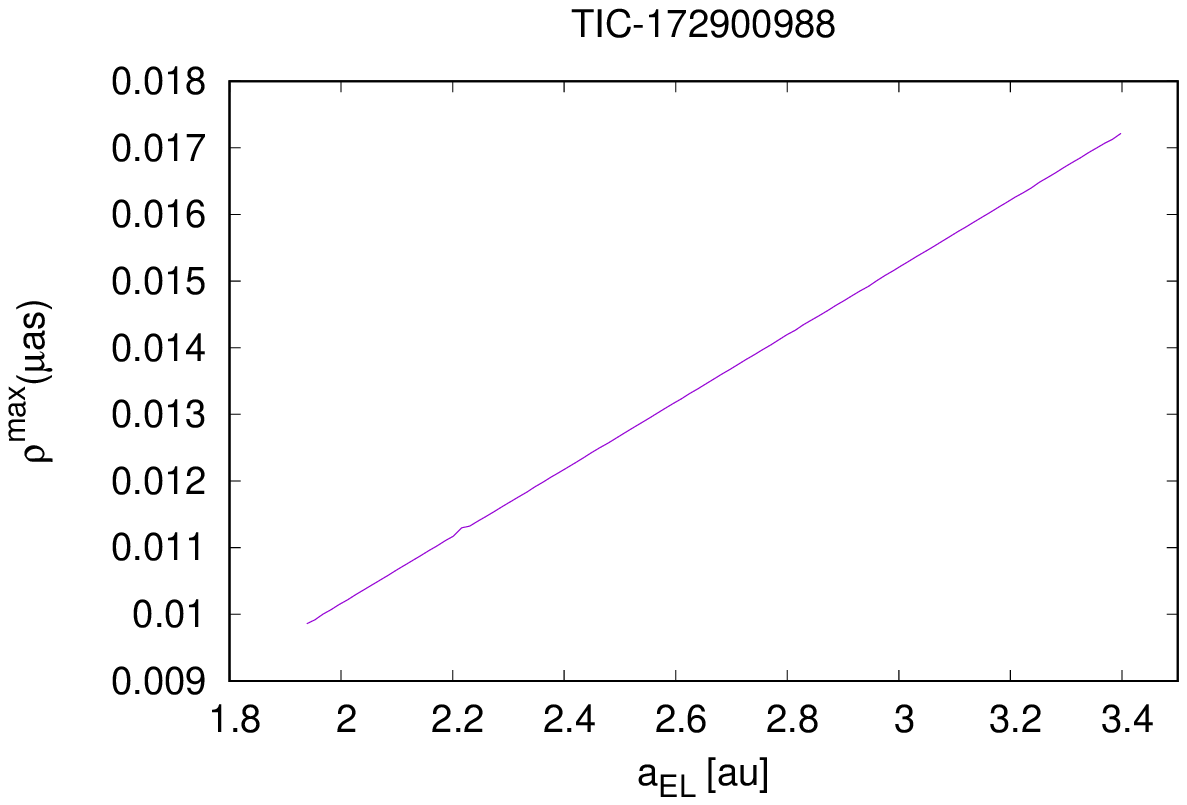}
\includegraphics[width=80mm,height=55mm]{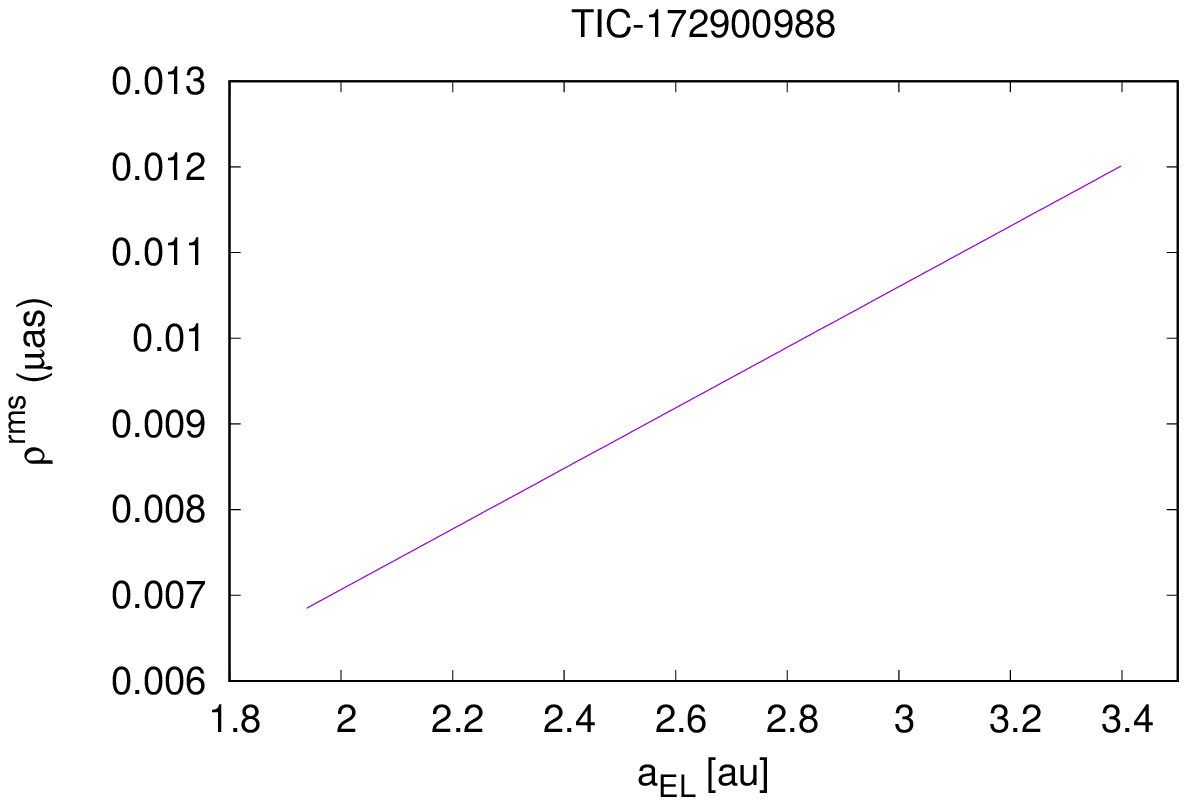}
\caption{Detectability values for TIC-172900988 (S4).}
\label{fig6}
\end{center}
\end{figure}

TIC-172900988 also displays small detection signals in its habitable zone.  The radial velocity signal is lower than the ones of TOI-1338, while the astrometric indicators are larger.  The total mass of the stellar binary is about twice as large as the one of TOI-1338. In addition, the inner edge of the habitable zone is about $50\%$ farther out than the inner border of
the habitable zone of TOI-1338. Finally, as it was mentioned previously, the maximum eccentricity acquired by the fictitious terrestrial planet through the habitable zone is smaller in the TIC-172900988 system.  On the other hand, the astrometric signals of TOI-1338 are smaller, mainly because of the fact that its distance $d$ from us is $399.017 pc$, which is much larger than the $246.263 pc$ of TIC-172900988. Fig. \ref{fig6} is a graphical representation of the radial velocity and astrometric signals for TIC-172900988.

\section{Summary and discussion}
We have explored the dynamics and potential habitability of the two circumbinary systems
that have been discovered with TESS so far. The binary system TOI-1338 hosts a planet that is about twice as massive as Neptune, while TIC-170900988 has a planet with a mass around three  Jupiter masses. We have explored the habitable zones of both planets in terms of their dynamical stability. We placed a fictitious Earth mass planet throughout the two habitable zones in order to check whether its orbit is dynamical stable and we carried out a number of numerical simulations over a sufficient amount of time.  From planetary formation theories it is possible to have such type of systems \citep[e.g.][]{2020MNRAS.494.1045B,2021MNRAS.504.6144B}.  We found that for the various orbital solutions of TOI-1338 and TIC-172900988 investigated, the Earth mass planetary orbit showed minimal variations in the semimajor axis and some small changes in the eccentricity. Hence, all configurations were found to be stable. Since the habitable zone of the system was found dynamically stable, we calculated dynamically informed habitable zones in order to get an assessment of the qualitative characteristics of the habitable zone.  We found in all cases that the permanently habitable zone covered a large part of the classical habitable zone.
In some cases that part was around $90\%$ of the latter, which means that the binary system is suitable of hosting low climate inertia terrestrial planets. This kind of result has been found in other circumbinary systems such as for example Kepler-38 or Kepler-64 \citep{2021FrASS...8...44G}. 

Subsequently, we checked what would be the magnitude of the detection signal for an
Earth mass planet in the habitable zone of the two binaries.  Using the radial velocity and astrometry detection methods, we made some estimates covering the whole range of the habitable zone. The results showed that considering our current detection facilities such discovery is not possible, especially if we try to use astrometry for detecting the planet.

\section*{Acknowledgements}
I would like to thank Siegfried Eggl who provided the code that
was used for our numerical experiments. I would also like to thank 
the referee O.C. Winter for his comments that helped me improve the
manuscript.

%%%%%%%%%%%%%%%%%%%%%%%%%%%%%%%%%%%%%%%%%%%%%%%%%%
\section*{Data Availability}

The data presented and discussed in this paper are available
upon reasonable request to the corresponding author.



\bibliographystyle{mnras}
\bibliography{ref} % if your bibtex file is called example.bib

%%%%%%%%%%%%%%%%%%%%%%%%%%%%%%%%%%%%%%%%%%%%%%%%%%

%%%%%%%%%%%%%%%%% APPENDICES %%%%%%%%%%%%%%%%%%%%%

\appendix

\section{Tables}
\onecolumn
\begin{table}
\caption{Physical parameters and orbital elements for TOI-1338 as given in \citet{2020AJ....159..253K} ({\it best solution}). $m$, $R$, $T_{eff}$, $P$, $a$, $e$, $\omega$ and $M$ denote mass, physical radius, effective temperature, orbital period, semimajor axis, eccentricity, argument of pericentre  and mean anomaly respectively.} 
\label{tab1}
\begin{center}	
\begin{tabular}{l l l l l l}
\hline\\
Stellar binary & &  &  &  & \\
\hline\\
$m_1=1.038 M_{\odot}$ & $m_2=0.2974 M_{\odot}$ & $R_1=1.299 R_{\odot}$ & $R_2=0.3015 R_{\odot}$ & $T_{eff1}=5990.7 K $ & $T_{eff2}=3317.1 K$\\
$P_b=14.608561 d$ & $a_b=0.1288 au$ & $e_b=0.15601$ & $\omega_b=117.561^{\circ}$ & $M_b=93.882^{\circ}$ & \\
\hline \\
Planet & & & & & \\
\hline \\
$m_p=30.2 M_{\oplus}$ & $P_p=95.141 d$ & $a_p=0.4491 au$ & $e_p=0.0928$ & $\omega_p=263.3^{\circ}$ & $M_p=128.3^{\circ}$ \\
       \hline
  \end{tabular}
\end{center}
\end{table}

\begin{table}
\caption{Physical parameters and orbital elements for TIC-172900988 as given in \citet{2021arXiv210508614K}. The symbols have the same meaning as in Table A1.} 
\label{tab2}
\begin{center}	
\begin{tabular}{l l l l l l}
\hline\\
Stellar binary & &  &  &  & \\
\hline\\
Solution 1 & &  &  &  & \\
$m_1=1.2382 M_{\odot}$ & $m_2=1.2025 M_{\odot}$ & $R_1=1.3844 R_{\odot}$ & $R_2=1.3169 R_{\odot}$ & $T_{eff1}=6050 K $ & $T_{eff2}=5983 K$\\
$P_b=19.657473 d$ & $a_b=0.191920 au$ & $e_b=0.44803$ & $\omega_b=69.603^{\circ}$ & $M_b=254.498^{\circ}$ & \\
Solution 2 & &  &  &  & \\
$m_1=1.2381 M_{\odot}$ & $m_2=1.2024 M_{\odot}$ & $R_1=1.3829 R_{\odot}$ & $R_2=1.3156 R_{\odot}$ & $T_{eff1}=6050 K $ & $T_{eff2}=5983 K$\\
$P_b=19.658073 d$ & $a_b=0.191920 au$ & $e_b=0.44806$ & $\omega_b=69.617^{\circ}$ & $M_b=254.491^{\circ}$ & \\
Solution 3 & &  &  &  & \\
$m_1=1.2382 M_{\odot}$ & $m_2=1.2025 M_{\odot}$ & $R_1=1.3833 R_{\odot}$ & $R_2=1.3150 R_{\odot}$ & $T_{eff1}=6050 K $ & $T_{eff2}=5983 K$\\
$P_b=19.658150 d$ & $a_b=0.191926 au$ & $e_b=0.44812$ & $\omega_b=69.618^{\circ}$ & $M_b=254.489^{\circ}$ & \\
Solution 4 & &  &  &  & \\
$m_1=1.2380 M_{\odot}$ & $m_2=1.2023 M_{\odot}$ & $R_1=1.3824 R_{\odot}$ & $R_2=1.3159 R_{\odot}$ & $T_{eff1}=6050 K $ & $T_{eff2}=5983 K$\\
$P_b=19.656660 d$ & $a_b=0.191907 au$ & $e_b=0.44822$ & $\omega_b=69.594^{\circ}$ & $M_b=254.504^{\circ}$ & \\
Solution 5 & &  &  &  & \\
$m_1=1.2377 M_{\odot}$ & $m_2=1.2021 M_{\odot}$ & $R_1=1.3827 R_{\odot}$ & $R_2=1.3141 R_{\odot}$ & $T_{eff1}=6050 K $ & $T_{eff2}=5983 K$\\
$P_b=19.658175 d$ & $a_b=0.191902 au$ & $e_b=0.44819$ & $\omega_b=69.621^{\circ}$ & $M_b=254.481^{\circ}$ & \\
Solution 6 & &  &  &  & \\
$m_1=1.2378 M_{\odot}$ & $m_2=1.2021 M_{\odot}$ & $R_1=1.3825 R_{\odot}$ & $R_2=1.3127 R_{\odot}$ & $T_{eff1}=6050 K $ & $T_{eff2}=5983 K$\\
$P_b=19.658014 d$ & $a_b=0.191905 au$ & $e_b=0.44814$ & $\omega_b=69.616^{\circ}$ & $M_b=254.487^{\circ}$ & \\
\hline \\
Planet & & & & & \\
\hline \\
Solution 1 & &  &  &  & \\
$m_p=822.4 M_{\oplus}$ & $P_p=188.763 d$ & $a_p=0.86733 au$ & $e_p=0.0890$ & $\omega_p=196.6^{\circ}$ & $M_p=143.8^{\circ}$ \\
Solution 2 & &  &  &  & \\
$m_p=844.1 M_{\oplus}$ & $P_p=190.388 d$ & $a_p=0.87231 au$ & $e_p=0.0664$ & $\omega_p=184.4^{\circ}$ & $M_p=328.1^{\circ}$ \\
Solution 3 & &  &  &  & \\
$m_p=835.3 M_{\oplus}$ & $P_p=193.995 d$ & $a_p=0.88333 au$ & $e_p=0.0767$ & $\omega_p=176.9^{\circ}$ & $M_p=276.7^{\circ}$ \\
Solution 4 & &  &  &  & \\
$m_p=869.1 M_{\oplus}$ & $P_p=199.001 d$ & $a_p=0.89843 au$ & $e_p=0.0904$ & $\omega_p=200.1^{\circ}$ & $M_p=13.9^{\circ}$ \\
Solution 5 & &  &  &  & \\
$m_p=942.1 M_{\oplus}$ & $P_p=200.452 d$ & $a_p=0.90274 au$ & $e_p=0.0271$ & $\omega_p=161.1^{\circ}$ & $M_p=157.1^{\circ}$ \\
Solution 6 & &  &  &  & \\
$m_p=981.0 M_{\oplus}$ & $P_p=204.051 d$ & $a_p=0.91355 au$ & $e_p=0.0265$ & $\omega_p=138.3^{\circ}$ & $M_p=11.8^{\circ}$ \\

       \hline
  \end{tabular}
\end{center}
\end{table}

%%%%%%%%%%%%%%%%%%%%%%%%%%%%%%%%%%%%%%%%%%%%%%%%%%

% Don't change these lines
\bsp	% typesetting comment
\label{lastpage}
\end{document}